\documentclass[12pt]{article}
\newcommand{\be}[3]{\begin{equation}  \label{#1#2#3}}     % non-hyper
\newcommand{\ee}{ \end{equation}}
\newcommand{\ba}{\begin{array}}
\newcommand{\ea}{\end{array}}
\newcommand{\NP}[3]{{\em Nucl. Phys.}{ \bf B#1#2#3}}

\renewcommand{\arraystretch}{1.8}
\begin{document}

\thispagestyle{empty}
\rightline{HUB-EP-97/012}
\rightline{UPR-736-T}
\rightline{hep-th/9702168}

\vspace{1truecm}

\centerline{\bf \large 
Subleading Contributions from Instanton Corrections }
\vspace{.3truecm}
\centerline{\bf \large 
in N = 2 Supersymmetric Black Hole Entropy
}
\vspace{1.2truecm}

\centerline{\bf Klaus Behrndt\footnote{e-mail: 
 behrndt@qft2.physik.hu-berlin.de}}
\vspace{.3truecm}
{\em
\centerline{Humboldt-Universit\"at, Institut f\"ur Physik}
\centerline{Invalidenstra\ss e 110, 10115 Berlin}
\centerline{Germany}
}
\vspace{1truecm}

\centerline{\bf 
Ingo Gaida\footnote{e-mail: gaida@cvetic.hep.upenn.edu}}
\vspace{.3truecm}
{\em
\centerline{Department of Physics and Astronomy}
\centerline{University of Pennsylvania}
\centerline{Philadelphia, PA 19104-6396, USA}
}

%%%%%%%%%%%%%%%%%%%%%%%%%%%%%%%%%%%%%%%%%%%%%%%%%%%%%%%%

\vspace{2.5truecm}

\begin{abstract}
We present subleading corrections to the $N=2$ supersymmetric black
hole entropy. These subleading contributions correspond to instanton
corrections of the Type II string theory. In particular we
consider an axion free black hole solution of low-energy effective Type
II string theory.  We present a procedure to include successiv all
instanton corrections.  Expanding these corrections at
particular points in moduli space yields polynomial and logarithmic
instanton corrections to the classical black hole entropy.  We comment
on a microscopic interpretation of these instanton corrections and
find that the logarithmic corrections correspond to subleading terms
in the degeneracy of the spectrum of an underlying quantum theory.
\end{abstract}

\bigskip \bigskip
\newpage

%%%%%%%%%%%%%%%%%%%%%%%%%%%%%%%%%%%%%%%%%%%%%%%%%%%%%%%%%%%%%%

\noindent
{\bf \large 1. Introduction} \bigskip 

\noindent
In recent times there has been substantial progress towards a better
understanding of the microscopic origin of the Bekenstein-Hawking
entropy. Using the $D$-brane approach a microscopic origin for certain
black hole solutions was found \cite{st/va}.  Using the fact that these
black holes are compactified string solution one can 
explore the known degeneracy formula of string states to find a
statistical interpretation of the Bekenstein-Hawking entropy.  In this
context Susskind speculated that the Bekenstein-Hawking entropy might
be explained in terms of the degeneracy of free strings \cite{su}.
For large level $N$ this degeneracy reads
\be002
d(N) = e^{S_{stat}} =  N^{-\frac{c+3}{4}}\ e^{2\pi\sqrt{\frac{c}{6}N}}  
\ee
where $c$ is the central charge. The exponential term is the leading
term while the polynomial part is known as the subleading term.
This picture has been recently reinvestigated and expanded in \cite{ho/po}.

\medskip

All supersymmetric black hole solutions with finite horizon in four
dimensions have a natural embedding as BPS states, breaking one half
of $N=2$ supersymmetry.  For all these solutions the leading exponential
term in (\ref{002}) could be identified at the classical level (torus
compactification) \cite{cv/hu} and if one includes quantum corrections
\cite{be/ca}.  On the other hand the origin of the subleading
polynomial terms in the context of the Bekenstein-Hawking entropy of
black hole solutions has up to now no analogous explanation.  The main
purpose of the present paper is to fill this gap.  We will restrict
ourselves to the entropy of BPS saturated Type II black hole
solutions.
In a forthcoming paper we will discuss the corresponding black hole
solutions of effective Type II and heterotic string theory and
the corresponding role of the subleading corrections in the context of
heterotic/Type II duality symmetries \cite{bg}.  For Schwarzschild
black holes the role of the subleading corrections has recently been
addressed in \cite{so}..

\medskip

BPS saturated black hole solutions become independent of the moduli on the
horizon \cite{Finn}.
This result can be understood from a
supersymmetric point of view \cite{fe/ka}: The central charge
appearing in the supersymmetry algebra reaches an extremum in moduli
space. The corresponding extremization problem is equivalent to the
solution of the equations ($p$ and $q$ are the magnetic and electric
charges)
\be010
Y^I - \bar{Y}^I = i p^I \qquad , \qquad F_I -\bar{F}_I = i q_I
\ee
with the symplectic vector
\be012
V = \pmatrix{Y^I \cr F_I}
\ee
which defines the special geometry of the theory ($F_I = \partial F/
\partial Y^I$ and $F$ is the prepotential) and the index $I= 0,1, .. n$ 
where $n$ is the number of vector multiplets. Note that in this
basis the symplectic constraint is: $i \langle \bar V , V \rangle =
|Z|^2$, where $Z$ is the $N=2$ central charge (for details see \cite{be/ca}).
The equations (\ref{010}) determine uniquely the $n+1$ fields $Y^I$
in terms of the charges and parameters of the prepotential, like
the intersection numbers, Euler  and instanton numbers. Taking
this solution the Bekenstein-Hawking entropy is
\be020
S = i \pi \; \left 
( \bar Y^{I}  F_{I}(Y) 
-  Y^{I} \bar F_{I} (\bar Y) 
\right )_{| extr.} \ = \  \pi \ |Z|_{|extr.}^{2}
\ee

\bigskip \bigskip

\noindent
{\large \bf 2. The Type II Solution}

\bigskip

\noindent
% In our procedure we will restrict ourselves mainly on the type II side.
We define the Kahler class moduli $t^A$ by
$t^A = Y^A / Y^0$ 
and the prepotential reads \cite{ca/os}
\be040
F(Y^0, Y^A) = (Y^0)^2 \left[ -{1\over 6} C_{ABC} t^A t^B t^C - 
{i \chi \zeta (3) \over
2 (2 \pi)^3} + {i \over (2 \pi)^3} \sum_{d_1,..d_n} n^r_{d_1, .. , d_n}
Li_3(e^{2 \pi i \, d_A t^A} ) \right].
\ee
Here $C_{ABC}$ are the classical intersection numbers, $\chi$ is the
Euler number, $n^r_{d_1, .. , d_n}$ are the numbers of genus zero
rational curves (instanton numbers) and $d_A$ are the degrees of these
curves. These numbers are determined by the Calabi-Yau threefold 
of the compactified effective Type II string theory.

Restricting ourselves to the axion free case ($t^A + \bar t^A =0$)
with $p^0 = q_A =0$ we find as a solution of (\ref{010})
\be050
Y^0 = {\lambda \over 2} \quad , \quad  Y^A = i\, { p^A \over 2} 
\qquad \mbox{and} \qquad t^A = i \, {p^A \over \lambda} 
\ee
and the  parameter $\lambda$ is fixed by the constraint $F_0 - \bar 
F_0 = iq_0$.  Inserting the solution into (\ref{040}) this constraint
becomes
\be060
4 {\partial \over \partial\lambda} F( \lambda , p^A) = i q_0 \ .
\ee
Note that for the axion free solution $F_0$ is pure imaginary
and $F_A$ is real. Inserting the fields in the entropy (\ref{020})
we obtain
\be070
S = \frac{\pi}{2} 
\left( 
- \lambda q_0 + {1 \over 2\lambda} C_{ABC} 
p^A p^B p^C  - {\lambda \over(2 \pi)^2} \; \sum_{d_1,..d_n} 
n^r_{d_1, .. , d_n} \;  d_A p^A  \; 
Li_2(e^{-2 \pi d_A p^A / \lambda} ) \right) \ .
\ee
With a given solution for $\lambda$ the entropy includes all instanton
corrections.  However, to find a well-defined entropy ($S \geq 0$) one
has to adjust the signs of the charges in an appropriate way.  At the
classical level and for the example we discuss below the entropy is
well-defined for large negative electric charge $q_0$.

\medskip

The entropy formula (\ref{070}) is implicit and it seems difficult to
find a general analytic solution.  However, we can expand the solution
at special points in moduli space.  Note first of all that for large
moduli ($t^A \gg 1$) the instanton part is exponentially suppressed
and the solution is determined by the pure cubic part.  Due to the
instanton nature it is impossible to expand around this point, since
we cannot obtain these terms in perturbation theory.  On the other
hand, if at least one of the moduli is small, say $t^3$, one finds a
non-trivial expansion of the instanton corrections near this special
point in moduli space. Thus, we consider the region
\be080
p^3 \ll \lambda \ll p^A
\ee
where $A \neq 3$. This limit corresponds to the case where one 4-cycle is
small, which is related to $t^3$, and all other 4-cycles are large. 
In this case the instanton correction reads
%\begin{eqnarray}
\be090
\sum_{d_A} n^r_{d_A} Li_3(e^{- 2 \pi d_A p^A / \lambda} ) 
%\approx
%\sum_{0,0,d_3,0,..} n^r_{0,0,d_3,0 ..} Li_3(e^{-2 \pi \, d_3 p^3 \over
%\lambda}) 
= \sum_{d_3} n^r_{d_3} \left(
\zeta(3) - {(2\pi)^3 d_3 p^3 \over 24\, \lambda} + ( {3 \over 4}
- {1 \over 2} \log{2\pi d_3 p^3 \over \lambda}) 
({2\pi d_3 p^3 \over \lambda})^2  \pm .. \right) \ ,
%\ea
\ee
with $n_{d_A}^r \equiv n_{d_1, .. ,d_n}^r$ and $n_{d_3}^r = 
n_{0,0,d_3, 0, ..}^r$. 
The logarithm corresponds to the non-analytic part of the trilogarithm. 
If we include further terms the structure remains
\be092
Li_3(e^{-x}) \approx  p(x) + q(x) \log x
\ee
where $p(x)$ and $q(x)$ are analytic polynomials for small $x$.
Using this expansion we find for the constraint (\ref{060}) 
\be100
q_0 \lambda^2 = - {1 \over 6} C_{ABC} p^A p^B p^C + 
{a\, \zeta(3) \over 2 (2\pi)^3} \lambda^3 - \sum_{d_3} 
n^r_{d_3} \left[{d_3 p^3 \over 24} \lambda^2 - {(d_3 p^3)^2 \over 
4\pi} \lambda\right] + {\cal O}((p^3)^3),
\ee
where the constant $a$ collects all terms proportional to $\zeta(3)$.
There is however a sublety in this expansion. Because there is no
restriction on the degrees of the rational curves the sum over all
instanton numbers ($n_{0,0,d_3,0, ..}^r$) can be ``in principle'' infinite.
% [However, calculations for the Calabi-Yau that is dual 
% to heterotic $STU$ model
% suggest that this sum can cut off ($d_{0,0,1}^r =-2$, $n_{0,0,d_3}^r=0$
% for $d_3=2, .. , 10$). BESSER]
However, calculations for the Calabi-Yau that is dual to the heterotic 
$STU$ model
indicate that this sum is finite ($n_{0,0,1}^r =-2$, $n_{0,0,d_3}^r=0$
for $d_3=2, .. , 10$). This is crucial for the correct mapping of both
theories in our approximation, see \cite{ca/cu} and the second ref.\ of
\cite{ca/os}. Therefore, we will assume that the sum converges. From
(\ref{080}) and (\ref{100}) follows, that $q_0 \gg p^A$ and we can expand
the solution as
\be110
\lambda = \sum_{i=1} {\alpha_i \over (\sqrt{q_0})^i} = \pm
\sqrt{- {1 \over 6} C_{ABC} p^A p^B p^C \over q_0} +
\sum_{d_3} n^r_{d_3} {(d_3 p^3)^{2} \over 8 \pi q_0} \pm .. \ .
\ee
In the end we considered two expansions.  First we used $p^3 \ll \lambda$
to obtain a power expansion in $p^3$. Then we expanded the solution of
(\ref{060}) in $1/\sqrt{q_0}$.  Finally, in order to obtain
the instanton corrections to the entropy, we have to expand
(\ref{070}) in the same way and insert this solution for $\lambda$. As
a result the entropy contains a polynomial part and logarithmic terms.
Taking only the leading terms for both parts into account (with the
minus sign for $\lambda$) the entropy reads
\be120
S = 2 \pi  \sqrt{- q_0 {1 \over 6} C_{ABC} p^A p^B p^C} -
\frac{1}{8}
\sum_{d_3} n^r_{d_3}  (d_3 p^3)^2 
\log 
\left(  
  \frac{ (d_3p^3)^{2} q_0}{ -\frac{1}{6}C_{ABC}p^A p^B p^C}
\right) \pm ..
\ee
or written in the exponential form
\be130
d = e^S =  e^{S_0} \ 
\prod_{d_3} 
\left( 
\frac{ (d_3 p^3)^{2} q_0 }{-\frac{1}{6}C_{ABC}p^A p^B p^C}
\right)^{-n^r_{d_3} (d_3 p^3)^{2}/8}
\ee 
with the classical entropy
\be140
S_0 = 2 \pi \sqrt{-q_0 {1 \over 6} C_{ABC} p^A p^B p^C}.
\ee
Choosing a special Calabi-Yau space this solution can be ``mapped on
the heterotic side'' \cite{theisen} and has an analogous
interpretation in a quantum corrected heterotic $N=2$ model
\cite{quantum}.  For the $STU$ model our expansion corresponds to the
heterotic prepotential
\be142
F  = i (Y^0)^2 \left[ STU \ - {1 \over \pi} (T-U)^2 \log(T-U) \ + \ 
\Delta(T,U) \right] \ .
\ee
Here the logarithmic and the polynomial corrections
$\Delta(T,U)$ correspond to one-loop corrections near the line
$T=U \neq 1$ \cite{quantum}.
If we take only the classical prepotential and the 
logarithmic quantum corrections into account, the entropy of the
corresponding quantum corrected axion free black hole in the $STU$ model 
with $N=|q_0|$ is 
\be144
 S  = \pi \ \left ( 
 \sqrt{N p^{1}p^{2}p^{3}} 
 - \frac{1}{2\pi} (p^{2}-p^{3})^{2} \ \mbox{log} \ \sqrt{N}
\right ).
\ee
Note that $p^1$ is an electric charge in the $STU$ model.

%%%%%%%%%%%%%%%%%%%%%%%%%%%%%%%%%%%%%%%%%%%%%%%%%%%%%%%%%%%%%%%%%%%%%%%%

\bigskip \bigskip

\noindent
{\large \bf 3. The Microscopic Picture}

\bigskip

\noindent
Up to now we have considered an instanton corrected black hole
solution of an effective string field theory in four dimensions.
However, the ultimate goal is to find a statistical interpretation of
the black hole entropy in terms of a degeneracy of states of an
underlying quantum theory.  At the classical level - without instanton
corrections - this has been established for the model under
consideration for the extremal and near-extremal case in \cite{be/mo}.
In this case the entropy is given by $S_0$ (\ref{140}) only.  The main
microscopic interpretation is as follows: The black hole carries one
electric and $n$ magnetic charges, corresponding to an electric
graviphoton and $n$ magnetic vector gauge fields.  There are two
different ways to consider the solution.  First, as a compactification
of the Type II string theory or, second, as a compactification of the
equivalent $M$-brane configuration.  The latter one is of special
interest. In this case one first compactifies the configuration down
to five dimensions on a Calabi-Yau manifold. In this case the 11d
configuration is an intersection of three 5-branes. These 5-branes
have to intesect over a common string \cite{ts}. After
compactification, this yields a magnetic string solution in 5d. Then
one wraps this magnetic string around the $5th$ direction. Thus, the
Bekenstein-Hawking entropy of the black hole is directly related to
the statistical entropy of a string. The corresponding black hole
states are momentum states of this magnetic string giving rise to the
electric charge in four dimensions. In this procedure one uses the
known formula for the degeneracy of string states, but one should keep
in mind, that one does not count states of the closed Type II
string. The momentum modes in 5d correspond to open membranes in 11d
which are attached to the $M$-5-branes.  All these states are located
at the intersection of the 5-branes, since they become massless there.
In a Calabi-Yau manifold the number of these intersections is given by
$C_{ABC}$. Furthermore, every intersection consists of many layers of
5-branes, which means that we have wrapped each of the 5-brane many
times around the different 4-cycles. This multiple wrapping is counted
by the magnetic charge $p^A$. Summing up all these contribution one
finds that the Bekenstein-Hawking entropy (\ref{140}) coincides in the
leading order (exponential term) with the statistical entropy of
string states
\begin{equation}
\label{ds}
d(N) = e^{S_{stat}} \sim N^{-\frac{r}{4}}\ e^{2\pi\sqrt{\frac{c}{6}N}}\ .
\end{equation}
Here $N=-q_0$ is the oscillator number level and $r$ is a parameter.
For a free string one has $r=c+3$ with $c= C_{ABC}p^A p^B p^C$ as the
effective central charge of the 5d string.

Our entropy formula (\ref{130}) indicate, that at least in parts the
polynomial term of this statistical entropy (\ref{ds}) can be
interpreted as instanton corrections.  What is the microscopic picture
of these terms?  The instanton terms correspond to mapped (euclidean)
closed string world sheets into the Calabi-Yau.  This mapping might be
of higher degree, e.g..\ multiple wrapping counted by the numbers
$d_A$.  Furthermore, the instanton numbers $n^r_{d_1, ... , d_n}$
count the number of rational curves upon which we map the string
worldsheet.  However, in $M$-theory we do not have string worldsheets
which we could wrap in the Calabi-Yau.  On the other hand one can find
the following interpretation in $M$-theory: Analogous to the
string worldsheet we have to wrap membranes completely into the
internal space. From the three (euclidean) worldvolume coordinates, one
is the $5th$ direction and the others are mapped onto the rational
curves.  These are additional winding states (in addition to the
momentum states discussed above).  Thus, in 11d our configuration
consists of i) the intersection of three 5-branes and ii) a gas of
free closed membranes.  On the Type II side the black hole is a compactified
intersection of three 4-branes and one 0-brane. This configuration
lives in a gas of free closed strings, which are mapped on rational
curves upon compactification. This string gas is the origin of the
subleading instanton corrections in the black hole entropy.

\medskip 

However, the occurrence of $\zeta(3)$ in (\ref{100}) is a serious
problem. In our approximation this term has been omitted.  In contrast
to all other parameters $\zeta(3)$ is not an integer or rational and
cannot be expressed in terms of $\pi$. As consequence it seems to be
difficult to find an appropriate microscopic interpretation.  An
interesting observation is that the statistical entropy of a massless
ideal gas for a membrane is proportional to $\zeta(3)$ \cite{kl/ts}:
$S_{stat.} = 7/8\pi \, \zeta(3)\,N \,L^2 \,T^{2}$ ($N$ number of free
bosons, $L^2$ is the spacial volume and $T$ is the tension).  It would
be very interesting to find a connection between this statistical
entropy and the $\zeta(3)$ correction to our Bekenstein-Hawking
entropy, which are caused by wrapped membranes. These $\zeta(3)$ terms
can be separated by considering a
region where all magnetic charges (which are related to 5-branes) are
neglected.  For this pure instantonic case we obtain
\be150
S = {(2 \pi)^4 (q_0)^2 \over 2 |a| \, \zeta(3)} \ .
\ee 
Turning off all 5-branes means that we have neglected all vector
multiplet contributions to the entropy except the one of the $N=2$
supergravity multiplet, which contains the graviphoton.  Thus, this
contribution has its origin in pure $N=2$ supergravity.

\bigskip \bigskip
\pagebreak

\noindent
{\large \bf 4. Summary}

\bigskip

\noindent
In this paper we have addressed the question of subleading
contributions to the entropy of $N=2$ supersymmetric
black holes. In low-energy effective
Type II theories in four dimensions these corrections have a possible
origin in instanton corrections.
% They give in the leading order additional polynomial as well as
% logarithmic corrections in the entropy (\ref{120}).
We have chosen particular points in moduli space to expand the
instanton corrections.  The calculations base on an expansion around a
vanishing 4-cycle in the Calabi-Yau space and the assumption that all
other 4-cycles are large.  In the context of the dual heterotic $STU$
model this expansion corresponds to the region $T\simeq U$ in moduli
space.  Since the corresponding black hole is a compactified string
solution in 5d, it is obvious to connect the Bekenstein-Hawking
entropy with the statistical entropy of an underlying quantum theory.
Since the formulas (\ref{002}) and (\ref{120}) have the same structure
we find that parts of the full instanton correction can serve as
subleading terms in the degeneracy formula of the underlying quantum
theory. They are corrections in a $1/\sqrt{N}$ ($N=|q_0|$) expansion
around a vanishing 4-cycle ($p^3 \simeq 0$).

%%%%%%%%%%%%%%%%%%%%%%%%%%%%%%%%%%%%%%%%%%%%%%%%%%%%%%%%%%%%%%%%%%%%%%%%%

\bigskip  \bigskip

\noindent
{\bf Acknowledgements}  \medskip \newline
We would like to thank Ed Derrick, Thomas Mohaupt and M. Cveti{\v{c}} 
for many fruitful discussions and comments.
The work of K.B.~is  supported by the DFG and the work of I.G.
~is supported by U.S. DOE Grant Nos. DOE-EY-76-02-3071 and the National 
Science Foundation Career Advancement Award No. PHY95-12732.

\renewcommand{\arraystretch}{1}

\end{document}